\documentclass[11pt]{article}

\usepackage{amssymb,amsthm, graphicx, stmaryrd, fancyhdr}
\usepackage{amsfonts}
\usepackage{amsmath}
\usepackage[english,german]{babel}
\usepackage{epsfig,epstopdf}
\usepackage{subfigure}
\usepackage{latexsym}
\usepackage{color}

\usepackage{amsmath,amssymb,epsfig, epstopdf}

\newtheorem{theorem}{Theorem}[section]
\newtheorem{definition}[theorem]{Definition}

\newtheorem{lemma}[theorem]{Lemma}

\newtheorem{remark}[theorem]{Remark}

\newtheorem{assumption}[theorem]{Assumption}

\def\R{\mathbb{R}}

\def\P{\mathbb{P}}

\def\R{\mathbb{R}}

\def\P{\mathbb{P}}

 \normalsize

\begin{document}
\selectlanguage{english}

 \title{Investment and Consumption  with Regime-Switching Discount Rates \thanks{  Work supported by NSERC grants 371653-09, 88051 and MITACS grants 5-26761, 30354 and the Natural Science Foundation of China (10901086).}}

\author{\normalsize    Traian A.~Pirvu \\[8pt]
        \small Dept of Mathematics \& Statistics\\
        \small McMaster University \\
        \small 1280 Main Street West \\
        \small Hamilton, ON, L8S 4K1\\
        \small tpirvu@math.mcmaster.ca
        \and
        \normalsize Huayue Zhang \\[8pt]
        \small Dept of Finance\\
        \small  Nankai University\\
        \small 94 Weijin Road \\
        \small Tianjin, China, 300071 \\
        \small  hyzhang69@nankai.edu.cn
\vspace*{0.8cm}}

\maketitle

\noindent {\bf Abstract.}
This paper considers the problem of consumption and investment in a financial market within
a continuous time stochastic economy. The investor exhibits a change in the discount rate. The investment opportunities are a stock and a riskless account. The market coefficients and discount factor switch according to a finite state Markov chain. The change in the discount rate leads to time inconsistencies of the investor's decisions. The randomness in our model is driven by a Brownian motion and a Markov
 chain. Following \cite{EkePir} we introduce and characterize the subgame perfect strategies. Numerical experiments show the effect of time preference on subgame perfect strategies and the pre-commitment strategies.
\vspace{1cm}


\begin{quote}

\end{quote}

\begin{flushleft}

\end{flushleft}

\setcounter{equation}{0}
\section{Introduction}

Dynamic asset allocation in a stochastic paradigm received
a lot of scrutiny lately. The pioneer works are \cite{Mer69} and \cite{Mer71}. Many works then followed, most
of them assuming an exponential discount function. \cite{EkePir} has given an overview of the literature in
the context of Merton portfolio management problem with exponential
discounting.

In this paper we consider a regime switching model for the financial market. This modeling is
consistent with some cyclicality observed in financial markets. Many papers consider these
types of markets for pricing derivative securities. Here we recall only two such works, \cite{Guo} and \cite{Eli}.
When it comes to optimal investment in regime switching markets we point to \cite{B}, \cite{Cad} and \cite{Z}. All these papers
 consider a constant rate of time preference.

 In our paper the discount rate is stochastic, exogenous and depends on the regime.
 By the best of our knowledge it is the first work to consider stochastic rates of time preference within the Merton problem framework.
 In a discrete time model \cite{Tosh} considers a cyclical discount factor. Next we motivate this modeling approach of discount
 rates. The issue of time discounting is the subject of many studies in financial economics.
Several papers stepped away from the exponential discounting modeling, and based
on empirical and experimental evidence proposed different discount models. 
In fact models with time varying discount rates have a long history, going back to
B\"{o}hm-Bawerk (1889) and Fisher (1930). More recent works like \cite{Bec} and \cite{Law}  
show that economic and social factors impact discount rates. 
\cite{Park} examins how the business cycles in the U.S. affects the rate of time preference.
Therefore it appears natural to assume that discount rates vary with the state of economy (so are regime
dependent).

Non constant discount rates lead to time inconsistency of the decision maker. 
The resolution is to consider subgame perfect strategies. These are
strategies which are optimal to implement now given that they will be implemented in the future. 
Ekeland and Lazrak \cite{EkeLaz} consider a deterministic problem with 
continuous time, namely the Ramsey problem of economic growth with non-exponential
 discounting. They define subgame perfect strategies and characterize them
 by a generalized HJB (the non local extension of HJB). Ekeland and Pirvu \cite{EkePir}
look at the Merton problem with special types of non-constant deterministic discount rates, and introduced/characterize
subgame perfect strategies. \cite{EkePirMbo} extends \cite{EkePir} by allowing general non-constant
deterministic discount rates and life insurance acquisition in their model. Related works are done
in \cite{Paz1}, \cite{Paz2} and \cite{Paz3}. \cite{Bjo} develops a general theory for stochastic control problems
 which are time inconsistent and introduced the subgame perfect subgame perefect tstrategies in a fairly general stochastic framework.
 Bj\"{o}rk, Murgoci and Zhou \cite{Bj2} look at
the mean variance problem with time changing risk aversion which is also time inconsistent.

In this paper, unlike previous works we consider a stochastic discount rates framework.
 The goal is to characterize the subgame perfect strategies.
 The methodology developed in  \cite{EkePirMbo} is employed to achieve this;
  it mixes the idea of value function (from the dynamic programming principle) with the idea that in the future ``optimal trading strategies'' are to be implemented (from the maximum principle of Pontryagin). The new twist in our paper is
 the Markov chain, and the mathematical ingredient used is It\^{o}'s formula for the Markov-modulated diffusions. Thus, we obtain a system  of four equations: first equation says that the value function is equal to the continuation utility of subgame perfect strategies; second equation is the wealth equation generated by subgame perfect strategies; the last two equations relate the value function to the subgame
 perfect strategies. The end result is a complicated system of PDEs, SDE and a nonlinear equation with
 a nonlocal term which can be solved for CRRA type utility. In this case  we find an ansatz for the value function ( by disentangling the time, the space and the Markov chain state component). This leads to subgame perfect strategies which are time/state dependent and linear in wealth.
 If constant discount rates, we notice that subgame perfect strategies coincide with the optimal ones. In a remark we discuss the pre-commitment
 strategies; these are strategies that a decision maker would implement if there exists a commitment mechanism. 
 
 Our contribution is to show that  stochastic rates of time preference
  lead to time inconsistent optimal strategies. Moreover, our resolution is to characterize subgame perfect strategies
  which are a substitute for the optimal strategies within this context.
 The contributions of this paper are: first, to consider a model with stochastic discount rates,
 and for the special case of CRRA preference to compare by numerical experiments 
  subgame perfect strategies and pre-commitment strategies. 
 
 Numerical findings show the dependence on the model's parameters of the subgame perfect strategies and pre-commitment strategies. 
 In one example we notice that a higher discount rate leads to a higher subgame perfect consumption rate; in another example we show that
 this is not the case if we allow the market parameters to switch.

The reminder of this paper is organized as follow.
 In section $2$ we describe the model and formulate the objective.
 Section $3$ contains the main result. Section 4 considers the special case of CRRA preferences. Section 5
 presents a numerical experiment. The paper ends with an appendix.

\section{The Model}

\subsection{The Financial Market}
Consider a probability space $(\Omega,\{\mathcal{F}_t\}_{0\leq t\leq
T},\mathcal{F},\P),$ which accommodates a standard Brownian
motion $W=\{W(t),{t\geq0}\}$ and a homogeneous finite state
continuous time Markov Chain (MC) $J=\{J(t), t\geq0\}.$ For simplicity assume that MC takes
values in $\mathcal{S}=\{\textbf{0},\textbf{1}\}.$ Our results hold true in the more general situation
of  $\mathcal{S}$ having finitely many states. The filtration $\{\mathcal{F}_t\}_{0\leq
t\leq \infty}$ is the completed
 filtration generated by $\{W(t)\}_{t\in[0,\infty)}$ and
 $\{J(t)\}_{t\in[0,\infty)}, $ that is $\mathcal{F}_t=\mathcal{F}_t^J
 \bigvee \mathcal{F}_t^W.$
  We assume that the stochastic processes
$W$ and $J$ are independent. The MC $J$ has a generator
$\Lambda=[\lambda_{ij}]_{\mathcal{S}\times\mathcal{S}}$ with
$\lambda_{ij}\geq0$\ for $i\neq j,$ and  $\sum_{j\in
\mathcal{S}}\lambda_{ij}=0$ for every $i\in \mathcal{S}.$ 

In our setup the financial market consists of a bank account $B$ and a
risky asset $S$, that are traded continuously over a finite time
horizon $[0,T]$ (here $T\in (0,\infty)$ is an exogenously given deterministic time).
 The price process of the bank account and risky asset are governed
 by the following Markov-modulated SDE:
 \begin{eqnarray}
 &&dB(t)=r(t,{J(t)})B(t)dt,\nonumber\\
&&dS(t)=S(t)\left[\alpha(t,{J(t)})\,dt
+\sigma(t,{J(t)})\,dW(t)\right],\quad0\leq t\leq \infty,\nonumber
 \end{eqnarray}
 where $B(0)=1$ and $S(0)=s>0$ are the initial prices. The functions   
 $r(t, i), \alpha(t,i), \sigma(t,i):  i\in \mathcal{S},$ are assumed to be
 deterministic, positive and continuous in $t.$  They represent the riskless rate, the stock return and the stock volatility (given the state $i$ of
 the MC at $t$). Moreover  
 $$\mu(t,i)\triangleq\alpha(t,i)-r(t,i) $$
 stands for the stock excess return.

\subsection{Investment-consumption strategies and wealth processes}
In our model, an investor continuously invests in the stock, bond and consumes. 
Let $\pi(t)$ be the dollar value invested in stock at time $t$ and
$c(t)\geq0$ be the consumption. ${X^{u}(t)}\triangleq X(t)$ represents the
wealth of the investor at time $t$ associated with the trading strategy $u=(\pi,c);$ 
it satisfies the following stochastic differential equation (SDE)
\begin{eqnarray}\label{equ:wealth-one}
dX^{u}(t)&=& \left(r(t,{J(t)}) X^{u}(t)+\mu(t,{J(t)})\pi(t)-c(t))\,dt
+\sigma(t,{J(t)})\pi(t)\,dW(t),\right.
\end{eqnarray}
where $X(0)=x>0$ is the initial wealth and $J_0=i\in{\mathcal{S}}$ is the initial state.
 This SDE is called the self-financing condition. An acceptable
investment-consumption strategy is defined below:
\begin{definition}
\label{def:portfolio-proportions} An $\R^{2}$-valued stochastic
process $\{u(t):=(\pi(t), c(t))\}_{t\in[0,\infty)}$ is called an
  admissible strategy process and write $u\in \mathcal{A}$ if
it is ${\mathcal{F}_t}-$ progressively measurable and it satisfies the following integrability condition
\begin{equation}%
\label{kj***} E[\int_0^t |\pi(s)\mu(s,J(s))-c(s)|\, ds+\int_0^t
      |\pi(s) \sigma({s,J(s)})|^2\, ds]<\infty, \text{ a.s., for all },
 t\in [0,\infty).
\end{equation}
and
\begin{equation}  \label{189}
{X}^{u}(t)>0,\,\, c(t)>0, \,t\in[0,T],\quad \mathbb{E} \!\!\!\!\sup_{\{t\leq s\leq T\}} |U(c(s))|<\infty,\,\,\mathbb{E}
|U({X}^{u}(T))|  <\infty.
\end{equation}
Here $U$ is a utility function and is defined in the next subsection.
\end{definition}
Under the regularity condition \eqref{kj***} imposed on $\{\pi(t),
c(t)\}_{t\in[0,\infty)}$ above, the SDE
 \eqref{equ:wealth-one} admits a unique strong solution.

 \subsection{ The risk preferences }
 A utility function $U$ is a strictly increasing, strictly concave differentiable real-valued function
defined on $[0,\infty)$ which satisfies the Inada conditions

\begin{equation}\label{in}
U'(0)\triangleq \lim_{x\downarrow 0} U'(x)=\infty,\qquad U'(\infty)\triangleq \lim_{x\rightarrow \infty} U'(x)=0.
\end{equation}

The strictly decreasing function $U'$ maps $(0,\infty)$ onto $(0,\infty)$ and hence has a strictly
decreasing inverse $I: (0,\infty)\rightarrow (0,\infty).$

\subsection{The discount rate}
As we mentioned in the introduction, this paper considers stochastic discount rates.
An easy way to achieve this is to let the discount rate depend on the state of the MC. Thus,
 at some intermediate time $t\in [0, T]$ the discount rate is $\rho_{J(t)},$ for some positive
 constants $\rho_{\bf{0}}$ and $\rho_{\bf{1}}.$ The intuition of this way of modeling discount
 rates stems from the connection between market states and discount rates (this can be explained
 by some models with endogenous discount rates which are influenced by economic factors).

\subsection{The Risk Criterion}
In our model, the investor decides what investment/consumption 
strategy to choose according to the expected utility risk criterion.
Thus, investor's goal is to maximize utility of intertemporal consumption
and final wealth. The novelty here is that we allow investor to update
the risk criterion and to reconsider the optimal strategies she/he computed in the past. 
This will lead to a time inconsistent behaviour as we show below. Let the agent start with a given
positive wealth $x,$ and a given market state $i,$ at some instant
$t.$ The $t-$optimal trading strategy
 $\{\tilde{\pi}_{t}({s}),\tilde{c}_{t}({s})\}_{s\in[t,T]}$ is chosen
to maximize the time $t$ optimization criterion
\begin{eqnarray}
\mathbb{E}\left[\int_{t}^{T}e^{-\rho_i(s-t)}
U(c(s))\,ds
+e^{-\rho_i(T-t)}{U}(X(T))|X(t)=x,J(t)=i\right].
\end{eqnarray}
 Throughout the paper we denote $\mathbb{E}_t^{x,i}\left[ \right] \triangleq \mathbb{E}\left[ |X(t)=x,J(t)=i\right].$
 The optimal $t-$trading strategy
 $\{\tilde{\pi}_{t}({s}),\tilde{c}_{t}({s})\}_{s\in[t,T]}$ is derived by the supermartingale/martingale principle.
 In the following $t$ stands for the initial time and $s$ stands for the time to go, i.e., $s\in[t,T].$ 
 Let $V(t,s,x,i)$ be
the value function associated with maximizing the time $t$ criterion
$$\sup_{u\in\mathcal{A}}\mathbb{E}\left[\int_{t}^{T}e^{-\rho_i(s-t)}
U(c(s))\,ds+e^{-\rho_i(T-t)}{U}(X(T))|X(t)=x,J(t)=i\right].$$
 This is characterized by the Hamilton-Jacobi-Bellman (HJB) equation
\begin{equation}\label{hjb}
\frac{\partial V}{\partial
s}(t,s,x,i)+\sup_{\pi,c}\left[(r x+\mu \pi-c) \frac{\partial
V}{\partial x}(t,s,x,i)+\frac{1}{2}\sigma^2\pi^{2}
\frac{\partial^{2} V}{\partial x^{2}}(t,s,x,i)+U(c)\right]\end{equation}$$-
\rho_iV(t,s,x,i)+\sum_{j\in \mathcal{S}}\lambda_{ij}V(t,s,x,j)=0,
$$
with the boundary condition
\begin{equation}\label{boundarycondition}
V(t,T,x,i)={U}(x).\ \ \ 
\end{equation}
Here $i$ stands for the value of MC at time $t.$ Thus, the HJB depends on the current time $t$ through $\rho_{J(t)}$ and this dependence is inherited by the $t-$optimal trading strategy. This in turn
leads to time inconsistencies. Let us detail on this; if at time 0 the market state is
\textbf{0}, that is $J(0)=\textbf{0}$, the HJB equation \eqref{hjb} depends on $\rho_{\textbf{0}},$
leading to  the corresponding time 0-optimal consumption strategy (which depends on $\rho_{\textbf{0}}$ through
the first order conditions) $\tilde{c}^{\rho_{\textbf{0}}}$; if at a later point in time $t_1$ the market state is \textbf{1}, that is $J(t_1)= \textbf{1}$, the HJB equation \eqref{hjb} depends on $\rho_{\textbf{1}}$, leading to  the corresponding
optimal consumption strategy (which depends on $\rho_{\textbf{1}}$ through the first order conditions) $\tilde{c}^{\rho_{\textbf{1}}}$; The time inconsistency is due to the fact that time 0-optimal strategy fails to remain optimal, i.e., $\tilde{c}^{\rho_{\textbf{0}}} (t)\neq \tilde{c}^{\rho_{\textbf{1}}}(t)$ for $t\geq t_1$. Notice that the value function $V$ depends on two time points $t$ and $s.$ In the case of constant rate of time preference $(\rho_{\textbf{0}}=\rho_{\textbf{1}}=\rho)$, the time 0 optimization criterion is
 $$\sup_{u\in\mathcal{A}} E\left[\int_0^Te^{-\rho s}
U({c}(s))ds +e^{-\rho T} U({X}(T)) |X(0)=x,J(0)=i\right],$$
and  the time $t$ optimization criterion is
$$\sup_{u\in\mathcal{A}} E\left[\int_t^Te^{-\rho (s-t)} U({c}(s))ds
+e^{-\rho (T-t)} U({X}(T)) |X(t)=x,J(t)=i \right].$$ 
Thus, the time 0-optimal strategy is also optimal for time $t$ optimization criterion.
This is no longer the case if $\rho_{\textbf{0}}\neq\rho_{\textbf{1}},$ and $J(0)\neq J(t).$

 One resolution of time inconsistency is to introduce subgame perfect strategies. They are optimal now given that they will be implemented in the future.

\section{The Main Result}

\subsection{The subgame perfect trading strategies}
For an admissible strategy $\{u(t)\triangleq{\pi}(t),{c}(t)\}_{t\in[0,T]}$
 and its corresponding wealth process
$\{X^{u}(t)\}_{t\in[0,T]}$ given by  \eqref{equ:wealth-one}, we denote
the expected utility functional by
\begin{equation}\label{01FUNCT}
\Theta(t,x,i,\pi,c)\triangleq\mathbb{E}_t^{x,i}\left[\int_{t}^{T}e^{-\rho_i(s-t)}
U(c(s)) \,ds+e^{-\rho_i(T-t)}{U}(X^u(T))\right].
\end{equation}
 Following \cite{EkePir} we shall give a rigorous mathematical
formulation of the subgame perefect strategies in the formal definition
below.

\begin{definition}\label{finiteh}
Let $F=(F_{1},F_{2}):[0,T]\times \R^+\times\mathcal{S}\rightarrow
{\R}^+\times\mathcal{S}$ be a map such that for any $t,x>0$ and $i\in
\mathcal{S}$
\begin{equation}\label{opt}
{\lim\inf_{\epsilon\downarrow 0}}\frac{ \Theta(t,x,i,F_{1},F_{2})-
\Theta(t,x,i,\pi_{\epsilon},c_{\epsilon})}{\epsilon}\geq 0,
\end{equation}
where
$$\Theta(t,x,i,F_{1},F_{2})\triangleq \Theta(t,x,i,\bar{\pi},\bar{c}),$$
\begin{equation}\label{0000eq}
\bar{\pi}(s)\triangleq{F_{1}(s,\bar{X}(s),J(s))},
\quad\bar{c}(s)\triangleq{F_{2}(s,\bar{X}(s),J(s))},
\end{equation}
and $\{\bar{\pi}(s),\bar{c}(s)\}_{s\in[t,T]}$ is admissible. Here, the  process $\{\bar{X}(s)\}_{s\in[t,T]}$ is the wealth corresponding to  $\{\bar{\pi}(s),\bar{c}(s)\}_{s\in[t,T]}.$ 
The process $\{{\pi}_{\epsilon}(s),{c}_{\epsilon}(s)\}_{s\in[t,T]}$
 is another admissible investment-consumption strategy defined by
\begin{equation}\label{1e}
\pi_{\epsilon}(s)=\begin{cases} \bar{\pi}(s),\quad
s\in[t,T]\backslash E_{\epsilon,t}\\
\pi(s), \quad s\in E_{\epsilon,t}, \end{cases}
\end{equation}

\begin{equation}\label{2e}
c_{\epsilon}(s)=\begin{cases} \bar{c}(s),\quad s\in[t,T]\backslash E_{\epsilon,t}\\
c(s), \quad s\in E_{\epsilon,t}, \end{cases}
\end{equation}
with $E_{\epsilon,t}=[t,t+\epsilon];$ $\{{\pi}(s),{c}(s)\}_{s\in
E_{\epsilon,t} }$ is any trading strategy for which
$\{{\pi}_{\epsilon}(s),{c}_{\epsilon} (s)\}_{s\in[t,T]}$ is an
admissible strategy. If \eqref{opt} holds true, then  $\{\bar{\pi}(s),\bar{c}(s)\}_{s\in[t,T]}$ is a subgame perfect strategy. 
\end{definition}

\subsection{The value function}
Our goal is in a first step to characterize the subgame perfect strategies and then to find them in special cases.  
Inspired by \cite{EkePir}, the value function $v:[0,T]\times \mathbb{R}^{+} \times\mathcal{S} \rightarrow \mathbb{R}$ is a $C^{1,2}$ function, concave in the second variable defined by
\begin{equation}\label{00ie1}
v(t,x,i)\triangleq\mathbb{E}_t^{x,i}\left[\int_{t}^{T}{e^{-\rho_i(s-t)}}
U(F_{2}(s,\bar{X}(s),J(s)))\,ds+
{e^{-\rho_i(T-t)}}{U}(\bar{X}(T))\right].
\end{equation}
 Recall that $\{\bar{X}(s)\}_{s\in[0,T]}$
is the wealth process  corresponding to  $\{\bar{\pi}(s),\bar{c}(s)\}_{s\in[t,T]},$ so it solves the SDE
\begin{eqnarray}\label{*10dyn}
d\bar{X}(s)&=&[r(s,J(s))\bar{X}(s)+\mu(s,J(s))
F_{1}(s,\bar{X}(s),J(s))-F_{2}(s,\bar{X}(s),J(s))]ds\\\notag&+& \sigma(s,J(s))
F_{1}(s,\bar{X}(s))dW(s).
\end{eqnarray}
 Moreover, $F=(F_{1},F_{2})$ is defined by
 \begin{equation}\label{109con}
 F_{1}(t,x,i)\triangleq-\frac{\mu(t,i)\frac{\partial v}{\partial x}(t,x,i)}
 {\sigma^2(t,i)
\frac{\partial^{2} v}{\partial x^{2}}(t,x,i)},\,\,
F_{2}(t,x,i)\triangleq\left(\frac{\partial v}{\partial
x}(t,x,i)\right),\,\,\,t\in[0,T],
\end{equation}
(recall that $I$ is the inverse marginal utility).
Thus, the value function is characterized by a system of four equations: one integral equation
with nonlocal term \eqref{00ie1}, one SDE \eqref{*10dyn} and two PDEs \eqref{109con}. Of course the existence of such a function $v$ satisfying the
equations above is not a trivial issue. 

\begin{assumption}\label{A2}
Assume that the PDE systems
\begin{eqnarray}\label{0PDEf}
&&\frac{\partial f}{\partial t}(t,s,x,i)+(r(t,i) x+\mu(t,i)
F_{1}(t,x,i)-F_{2}(t,x,i))\frac{\partial f}{\partial
x}(t,s,x,i)\nonumber\\&+&\!\!\!\!\!\frac{\sigma^2(t,i) F_{1}^{2}(t,x,i)}{2}
\frac{\partial^2 f}{\partial x^2}(t,s,x,i)\!\!+\!\!\sum_{j\in
\mathcal{S}}\lambda_{ij}f(t,s,x,j)=0, f(t,t,x,i)=U(F_2(t,x,i))
\end{eqnarray}

\begin{eqnarray}\label{0PDE1}
&&\frac{\partial h}{\partial t}(t,x,i)+(r(t,i) x+\mu(t,i)
F_{1}(t,x,i)-F_{2}(t,x,i))\frac{\partial h}{\partial
x}(t,x,i)\nonumber\\&+&\frac{\sigma^2(t,i) F_{1}^{2}(t,x,i)}{2}
\frac{\partial^2 h}{\partial x^2}(t,x,i)+\sum_{j\in
\mathcal{S}}\lambda_{ij}h(t,x,j)=0,\,\,h(T,x,i)=U(x).
\end{eqnarray}
have a $C^{1,2}$ solution on $[0,s]\times \mathbb{R}^{+}\times\mathcal{S}\rightarrow \mathbb{R}$ with exponential growth. Here $t<s\leq T,$
and $i\in\mathcal{S}.$ Moreover assume that the SDE \eqref{*10dyn} has a solution and that
$\{\bar{\pi}(s),\bar{c}(s)\}_{s\in[0,T]}$ given by \eqref{0000eq} with $(F_1, F_2)$ of \eqref{109con} is admissible.\end{assumption}

In the light of this assumption and by the Feyman-Kac Theorem it follows that 
$$f(t,s,x,i)=\mathbb{E}_t^{x,i}[U(F_{2}(s,\bar{X}(s),J(s)))],\quad h(t,x,i)=\mathbb{E}_t^{x,i}[U(\bar{X}(T))].$$
By \eqref{00ie1}
\begin{equation}\label{01ie1}
v(t,x,i)=\int_{t}^{T}e^{-\rho_i(s-t)} f(t,s,x,i)\,ds +
e^{-\rho_i(T-t)} h(t,x,i).
\end{equation}
Let us also define
\begin{equation}\label{001ie1}
\bar{v}(t,x,i)\triangleq\int_{t}^{T}e^{-\rho_j(s-t)} f(t,s,x,i)\,ds +
e^{-\rho_j(T-t)} h(t,x,i),\,i,j\in\mathcal{S}. 
\end{equation}
The next Lemma expresses the integral equation \eqref{00ie1} as a system of PDEs. 

\begin{lemma}\label{lala}
The value function $v$ solves the following system of PDEs (four equations)
\begin{equation}\label{zhang}
\frac{\partial v}{\partial t}(t,x,i)+(r(t,i) x+\mu(t,i) F_{1}(t,x,i)-F_{2}(t,x,i))
\frac{\partial v}{\partial x}(t,x,i) +\frac{\sigma^2(t,i)
F_{1}^{2}(t,x,i)}{2} \frac{\partial^2 v}{\partial x^2}(t,x,i)\end{equation}
$$+U(F_{2}(t,x,i))+\lambda_{ii}v(t,x,i)-\rho_i v(t,x,i)=-\lambda_{ij}\bar{v}(t,x,j),\,i,j\in\mathcal{S}. 
$$

\begin{equation}\label{zhang1}\frac{\partial \bar{v}}{\partial t}(t,x,i)+(r(t,i) x+\mu(t,i) F_{1}(t,x,i)-F_{2}(t,x,i))
\frac{\partial \bar{v}}{\partial x}(t,x,i) +\frac{\sigma^2(t,i)
F_{1}^{2}(t,x,i)}{2} \frac{\partial^2 \bar{v}}{\partial x^2}(t,x,i)\end{equation}
$$+U(F_{2}(t,x,i))+\lambda_{ii}\bar{v}(t,x,i)-\rho_j \bar{v}(t,x,i)=-\lambda_{ij}{v}(t,x,j),\,i,j\in\mathcal{S}. 
$$  with the boundary condition
$v(T,x,i)=\bar{v}(T,x,i)= U(x).$

Conversely if there exists a $C^{1,2}$ solution $(v, \bar{v})$ on $[0,T]\times \mathbb{R}^{+}\times\mathcal{S}\rightarrow \mathbb{R}$
with exponential growth for \eqref{zhang}, \eqref{zhang1}, then $v$ satisfies \eqref{00ie1}. 
\end{lemma}

\noindent Proof: By differentiating \eqref{01ie1} with respect to $t$ we get

\begin{equation}\label{0t1ie1}
\frac{\partial v}{\partial t}(t,x,i)=\int_{t}^{T}e^{-\rho_i(s-t)}
\frac{\partial f}{\partial t}(t,s,x,i)\,ds+ e^{-\rho_i(T-t)}
\frac{\partial h}{\partial t}(t,x,i)+\rho_i v(t,x,i)-f(t,t,x,i).
\end{equation}
Moreover
\begin{equation}\label{0xie1}
\frac{\partial v}{\partial x}(t,x,i)=\int_{t}^{T}e^{-\rho_i(s-t)}
\frac{\partial f}{\partial x}(t,s,x,i)\,ds
+e^{-\rho_i(T-t)}\frac{\partial h}{\partial x}(t,x,i).
\end{equation}

\begin{equation}\label{0yie1}
\frac{\partial^2 v}{\partial x^2}(t,x)=\int_{t}^{T}e^{-\rho_i(s-t)}
\frac{\partial^2 f}{\partial x^2}(t,s,x,i)\,ds
+e^{-\rho_i(T-t)}\frac{\partial^2 h}{\partial x^2}(t,x,i).
\end{equation}
In light of \eqref{0PDEf}, \eqref{0PDE1}, \eqref{0t1ie1},
\eqref{0xie1} and \eqref{0yie1} it follows that

$$\frac{\partial v}{\partial t}(t,x,i)+(r(t,i) x+\mu(t,i) F_{1}(t,x,i)-F_{2}(t,x,i))
\frac{\partial v}{\partial x}(t,x,i) +\frac{\sigma^2(t,i)
F_{1}^{2}(t,x,i)}{2} \frac{\partial^2 v}{\partial x^2}(t,x,i)$$
$$+U(F_{2}(t,x,i))+\lambda_{ii}v(t,x,i)-\rho_i v(t,x,i)=-\lambda_{ij}\bar{v}(t,x,j).
$$
Similarly we get the PDE for $\bar{v}.$
For the converse let us define
$$V(t,x,i,\rho)\triangleq\mathbb{E}_t^{x,i}\left[\int_{t}^{T}{e^{-\rho(s-t)}}
U(F_{2}(s,\bar{X}(s),J(s)))\,ds+
{e^{-\rho(T-t)}}{U}(\bar{X}(T))\right]. $$
Then by Feyman-Kac Theorem it follows that 
$$v(t,x,i)=V(t,x,i,\rho_i),\quad \bar{v}(t,x,i)=V(t,x,i,\rho_j),\,i,j\in\mathcal{S}, $$
so $v$ satisfies \eqref{00ie1}. 
\begin{flushright}
$\square$
\end{flushright}
The following Theorem states the central result of our paper.
\begin{theorem}\label{main0}
Under Assumption \ref{A2} the trading strategy $\{\bar{\pi}(s),\bar{c}(s)\}_{s\in[0,T]}$ given by \eqref{0000eq}
with $(F_1, F_2)$ of \eqref{109con} is a  subgame perfect strategy. 
\end{theorem}
\noindent Proof: Let us define

\begin{eqnarray}
\overline{\Gamma} v(t,x,i)&\triangleq&\frac{\partial v}{\partial
t}(t,x,i)+\left(r(t,i)x-I\left(\frac{\partial v}{\partial
x}(t,x,i)\right)\right)\frac{\partial v}{\partial
x}(t,x,i)\nonumber\\&-&
\frac{\mu^{2}(t,i)}{2\sigma^{2}(t,i)}\frac{{[\frac{\partial
v}{\partial x}}(t,x,i)]^{2}}{\frac{\partial^{2} v}{\partial
x^{2}}(t,x,i)}+\sum_{j\in\mathcal{S}}\lambda_{ij}(v(t,x,i)-v(t,x,j))+U(F_{2}(t,x,i)).
\end{eqnarray}

\begin{eqnarray}
{\Gamma}^{\pi, c} v(t,x,i)&\triangleq&\frac{\partial v}{\partial
t}(t,x,i)+\left(r(t,i)x-\mu(t,i)\pi-c\right)\frac{\partial
v}{\partial x}(t,x,i)\nonumber\\&+&\frac{1}{2}
{\sigma^{2}(t,i)}\pi^2{\frac{\partial^{2} v}{\partial
x^{2}}(t,x,i)}+\sum_{j\in\mathcal{S}}\lambda_{ij}(v(t,x,i)-v(t,x,j))+U(c).
\end{eqnarray}
By \eqref{zhang}
$$ \overline{\Gamma} v(t,x,i)=\lambda_{ij}(v(t,x,j)-\bar{v}(t,x,j)),\,i,j\in\mathcal{S}. $$
The concavity of $v$ and the first order conditions lead to
\begin{equation}\label{021}
\overline{\Gamma} v(t,x,i)=\max_{\{\pi,c \}} {\Gamma}^{\pi, c} v(t,x,i),\qquad (F_{1} (t,x,i), F_{2} (t,x,i))=\arg\max_{\{\pi,c \}}{\Gamma}^{\pi, c} v(t,x,i).
\end{equation}

Let us recall that
\begin{equation}\label{yy}
\Theta(t,x,i,F_{1},F_{2})=v(t,x,i).
\end{equation}
Thus
\begin{eqnarray}\label{7.16}
&&\Theta(t,x,i,F_{1},F_{2})-\Theta(t,x,i,\pi_{\epsilon},c_{\epsilon})\nonumber\\
&=&\mathbb{E}_t^{x,i}\left[\int_{t}^{t+\epsilon}e^{-\rho_i(s-t)}
[U(F_2(s,\bar{X}(s),J(s)))-U(c(s))]\,ds\right]\nonumber\\
&+&\mathbb{E}_t^{x,i}\left[\int_{t+\epsilon}^Te^{-\rho_i(s-t)}
[U(F_2(s,\bar{X}(s),J(s)))-U(c(s))]\,ds\right]\nonumber\\
&+&\mathbb{E}_t^{x,i}\left[  e^{-\rho_i(T-t)}(U(\bar{X}(T))-U({X}(T)))\right].
\end{eqnarray}
In the light of inequalities \eqref{189} and
Dominated Convergence Theorem
\begin{eqnarray}
{\lim_{\epsilon\downarrow 0}}
\frac{\mathbb{E}_t^{x,i}\left[\int_{t}^{t+\epsilon}e^{-\rho_i(s-t)}
[U(F_2(s,\bar{X}(s),J(s)))-U(c(s))]\,ds\right]}{\epsilon}=
U(F_2(t,x,i)-U(c(t)).\nonumber \end{eqnarray}
By \eqref{yy} it follows that
\begin{eqnarray}\label{7.19}
&&\mathbb{E}_t^{x,i}\left[\int_{t+\epsilon}^Te^{-\rho_i(s-t)}
[U(F_2(s,\bar{X}(s),J(s)))-U(c(s))]\,ds\right]\\\nonumber&+&\mathbb{E}_t^{x,i}
\left[e^{-\rho_i(T-t)}(U(\bar{X}(T))-U({X}(T)))\right]\nonumber\\
&=&\mathbb{E}_t^{x,i}\left
[v(t+\epsilon,\bar{X}(t+\epsilon),J(t+\epsilon))
-v(t+\epsilon,{X}(t+\epsilon),J(t+\epsilon))\right]\nonumber\\
&+&\mathbb{E}_t^{x,i}\left[\mathbb{E}[\int_{t+\epsilon}^T
(e^{-\rho_i(s-t)}-e^{-\rho_{J(t+\varepsilon)}(s-t)})
[U(F_2(s,\bar{X}(s),J(s)))-U(F_2(s,{X}(s),J(s)))]\,ds\right]\nonumber\\
&+&\mathbb{E}_t^{x,i}\left[\mathbb{E}[
(e^{-\rho_i(T-t)}-e^{-\rho_{J(t+\varepsilon)}(T-t+\epsilon)})
[U(\bar{X}(T))-U({X}(T))]\right].
\end{eqnarray}

By inequalities \eqref{189} and
Dominated Convergence Theorem it follows that

$${\lim_{\epsilon\downarrow 0}}\frac{\mathbb{E}_t^{x,i}\left[
(e^{-\rho_i(T-t)}-e^{-\rho_{J(t+\varepsilon)}(T-t+\epsilon)})
[U(\bar{X}(T))-U({X}(T))]\right]}{\varepsilon}$$

$$=\lambda_{ij}
(e^{-\rho_i(T-t)}-e^{-\rho_{j}(T-t)})\mathbb{E}_t^{x,i}\left[
[U(\bar{X}(T))-U({X}(T))]\right]=0.$$
By the same token one can get that
$${\lim_{\epsilon\downarrow 0}}\frac{
\mathbb{E}_t^{x,i}\left[\int_{t+\epsilon}^T
(e^{-\rho_i(s-t)}-e^{-\rho_{J(t+\varepsilon)}(s-t)})
[U(F_2(s,\bar{X}(s),J(s)))-U(F_2(s,{X}(s),J(s)))]\,ds\right]}{\epsilon}=0.$$
It\^{o}'s formula yields
\begin{eqnarray*}
&&\mathbb{E}_t^{x,i}\left[v(t+\epsilon,\bar{X}(t+\epsilon),J(t+\epsilon))
-v(t+\epsilon,{X}(t+\epsilon),J(t+\epsilon))\right]\nonumber\\
&=&\mathbb{E}_t^{x,i}\int_t^{t+\epsilon}[\overline{\Gamma}
v(s,\bar{X}(s),J(s))- U(F_{2}(s,\bar{X}(s),J(s))]ds -\mathbb{E}_t^{x,i}\int_t^{t+\epsilon}[\Gamma^{\pi,c}
v(s,{X}(s),J(s))-U(c(s))]ds,
\end{eqnarray*}
Therefore
\begin{eqnarray}
{\lim_{\epsilon\downarrow 0}}\frac{\Theta(t,x,i,F_{1},F_{2})-\Theta(t,x,i,
\pi_{\epsilon},c_{\epsilon})}{\epsilon}
= [\overline{\Gamma} v(t,x,i)-{\Gamma}^{\pi,c} v(t,x,i)]\geq0,\nonumber
\end{eqnarray}
(the inequality follows from \eqref{021}).

\begin{flushright}
$\square$
\end{flushright}

\section{CRRA Preferences}
In this section we assume that the utility is of power type,
 i.e., $U(x)=U_{\gamma}(x)=\frac{x^\gamma}{\gamma}.$
Let us take advantage of the special form of the utility function to simplify the
problem of finding $v.$ When $\gamma\neq 0$ we search for $v$ of the form: \begin{equation}\label{oP22}
  v(t,x,i)=g(t,i)\frac{x^\gamma}{\gamma},\ \ x\geq0.
\end{equation}
When $\gamma=0$ (logarithmic utility) 
we look for $v$ of the form
\begin{equation}\label{loP22}
  v(t,x,i)=g(t,i)\log{x}+l(t,i),\ \ x\geq0.
\end{equation}
The functions $g(t,i)$ and $l(t,i)$ are to be found.
 In light of equations (\ref{109con}) one gets
\begin{equation}\label{eQ1}
 F_1(t,x,i)=\frac{\mu(t,i) x}{\sigma^2(t,i) (1-\gamma)},\ \
 F_2(t,x,i)={g^{\frac{1}{\gamma-1}}(t,i)}x.
\end{equation}
Note that the function $l(t,i)$ does not enter the above equation
so its expression is not important. By (\ref{*10dyn}), the associated wealth process satisfies  the
following SDE:
\begin{eqnarray}\label{dyn1}
d\bar{X}(s)&=&\left[r(s,J(s))+\frac{\mu^2(s,J(s))}{\sigma^2(s,J(s))(1-\gamma)}
-g^{\frac{1}{\gamma-1}}(s,J(s))\right]\bar{X}(s)ds\nonumber\\
&&+\frac{\mu(s,J(s))}{\sigma(s,J(s))(1-\gamma)}\bar{X}(s)dW(s).
\end{eqnarray}
This is a linear SDE which can be easily solved. By plugging $v$ of (\ref{oP22}) into
(\ref{00ie1}) (with $F_1,F_2$ of (\ref{eQ1}) and $\bar{X}$ of (\ref{dyn1}), we obtain the following system of four equations for
$g(t,i), \bar{g}(t,i),\,\, i\in \mathcal{S}:$
\begin{equation}\label{oDe} \frac{\partial g}{\partial t}(t,i)+[\gamma r(t,i)
+\frac{\mu^2(t,i)\gamma}{2\sigma^2(t,i)(1-\gamma)}-\rho_i]g(t,i)+
\lambda_{ii}g(t,i)+(1-\gamma)g^{\frac{\gamma}{\gamma-1}}(t,i)=-\lambda_{ij}\bar{g}(t,j),
\end{equation}
\begin{equation}\label{pir}
\frac{\partial \bar{g}}{\partial t}(t,i)+[\gamma r(t,i)
+\frac{\mu^2(t,i)\gamma}{2\sigma^2(t,i)(1-\gamma)}-\rho_j]\bar{g}(t,i)+
\lambda_{ii}\bar{g}(t,i)+(1-\gamma)g^{\frac{1}{\gamma-1}}(t,i)\bar{g}(t,i)=-\lambda_{ij}{g}(t,j),
\end{equation} with the
final condition $g(T,i)=\bar{g}(T,i)=1,\,i\in \mathcal{S}.$ Next we show that there exists a unique solution for this ODE system. Let us summarize  \begin{lemma}\label{l4.1} There exists a unique continuously differentiable uniformly bounded
 solution \\$g(t,i), \bar{g}(t,i),\, i\in \mathcal{S}$ for the system (\ref{oDe}) and (\ref{pir}). Furthermore, $v(t,x,i)=g(t,i)\frac{x^\gamma}{\gamma}$ is a value function, meaning that $v$ is continuously differentiable, concave in $x$ and satisfies \eqref{00ie1} with $F_1, F_2$ of
(\ref{eQ1}) and $\bar{X}$ of (\ref{dyn1}).\end{lemma}
Appendix A. proves this Lemma.
\begin{flushright}
$\square$
\end{flushright}


\begin{theorem}\label{4.1}
 The trading strategy $\{\bar{\pi}(s),\bar{c}(s)\}_{s\in[0,T]}$ given by 
 \begin{equation}\label{eQ}
 \bar{\pi}(s)=\frac{\mu(s,J(s)) \bar{X}(s)}{\sigma^2(s,J(s)) (1-\gamma)},\ \
\bar{c}(s)={g^{\frac{1}{\gamma-1}}(s,J(s))} {\bar{X}(s)},
\end{equation}
is a subgame perfect strategy.
\end{theorem}
\noindent Proof: Let us notice that the boundedness of $g$ and the fact that
$\bar{X}$ of \eqref{dyn1} is positive and it has finite moments of any order imply the acceptability
of  $\{\bar{\pi}(s),\bar{c}(s)\}_{s\in[0,T]}.$ Next, the conclusion follows from Theorem \ref{main0}.
\begin{flushright}
$\square$
\end{flushright}

\begin{remark}
In the case of constant discount rate, i.e., $\rho_{\bf{0}}=\rho_{\bf{1}},$ the subgame
perfect strategies coincide with the optimal ones. This can be seen by looking at the
equation \eqref{zhang} which is exactly HJB \eqref{hjb}
(after the first order conditions are implemented and by using $v=\bar{v}$ (see \eqref{01ie1} and\eqref{001ie1} ).
 The optimal investment and consumption problem is solved in closed form by \cite{Cad} within this framework.
\end{remark}

\begin{remark}
The subgame perefect investment strategy does not change relative to the Merton model with constant discount rate;
more precisely the proportion of wealth invested in the stocks is the same as in the case of Merton model
with constant discount rate. This can be explained by the randomness of market coefficients being driven by
the Markov chain only. We conjecture that in a model with mean reverting
market price of risk this result does not hold.
\end{remark}

\begin{remark}\label{ab}
There are three important trading strategies:
\begin{itemize}
 \item the subgame perfect strategy $\{\bar{\pi}(s),\bar{c}(s)\}_{s\in[0,T]}$ (with associated wealth process $ \{\bar{X}(s)\}_{s\in[0,T]}$ ) given by
  $$\bar{\pi}(s)=\frac{\mu(s,J(s)) \bar{X}(s)}{\sigma^2(s,J(s)) (1-\gamma)},\ \
\bar{c}(s)={g^{\frac{1}{\gamma-1}}(s,J(s))} {\bar{X}(s)},$$
\item two pre-commitment strategies corresponding to the two discount rates $\rho_{ \bf{0} }, \rho_{ \bf{1} }$ (with associated wealth process $ \{\hat{X}_{k}(s)\}_{s\in[0,T]},\, k=1,2$  ) which are optimal
at time 0, but fail to remain optimal afterwards; they are given by
 $$\hat{\pi}_{k}(s)=\frac{\mu(s,J(s)) \hat{X}_{k}(s)}{\sigma^2(s,J(s)) (1-\gamma)},\ \
\hat{c}_{k}(s)={\hat{g}^{\frac{1}{\gamma-1}}_{k}(s,J(s))} {\hat{X}_{k}(s)}, \,\,k=1,2;$$ 
here $\hat{g}_{k}(t,i),\, i,j\in \mathcal{S},\,k=1,2$ solve:
\begin{equation}\label{aaa}
 \frac{\partial \hat{g}_{1}}{\partial t}(t,i)+[\gamma r(t,i)
+\frac{\mu^2(t,i)\gamma}{2\sigma^2(t,i)(1-\gamma)}-\rho_{ \bf{0} }]\hat{g}_{1}(t,i)+
\lambda_{ii}\hat{g}_{1}(t,i)+(1-\gamma)\hat{g}_{1}^{\frac{\gamma}{\gamma-1}}(t,i)=-\lambda_{ij}\hat{g}_{1}(t,j),
\end{equation}
\begin{equation}\label{aaa1}
\frac{\partial \hat{g}_{2}}{\partial t}(t,i)+[\gamma r(t,i)
+\frac{\mu^2(t,i)\gamma}{2\sigma^2(t,i)(1-\gamma)}-\rho_{ \bf{1} }]\hat{g}_{2}(t,i)+\lambda_{ii}\hat{g}_{2}(t,i)+(1-\gamma)\hat{g}_{2}^{\frac{\gamma}{\gamma-1}}(t,i)=-\lambda_{ij}\hat{g}_{2}(t,j),
\end{equation}
with boundary condition $\hat{g}_{1}(T,i)=\hat{g}_{1}(T,i), \, i\in \mathcal{S}.$
\item a naive strategy that switches in between the two pre-commitment strategies.
\end{itemize}
 \end{remark}

\section{Numerical Analysis }

In this section, we use Matlab's powerful ODE solvers (especially the functions  ode23 and ode45) to perform
numerical experiments. We numerically solve ODE systems \eqref{oDe}, \eqref{aaa} and \eqref{aaa1} to get the subgame
perfect  and pre-commitment strategies.  We take the Markov Chain generator to be
\begin{equation}
\left(
\begin{array}{cc}
 -2 & 2 \\
 1.5 & -1.5
 \end{array}
\right)\nonumber
\end{equation}
We plot the subgame perfect and pre-commitment consumption rates denoted by 
$$\bar{C}(t,J(t))\triangleq\frac{F_2(t,\bar{X}(t),J(t))}{\bar{X}(t)}=g^{\frac{1}{\gamma-1}}(t,J(t)),$$
$$\hat{C}_{k}(t,J(t))\triangleq\frac{\hat{c}_{k}(t)}{\hat{X}_{k}(t)}=\hat{g}_{k}^{\frac{1}{\gamma-1}}(t,J(t)),\,\, k=1,2.$$


\noindent Fig. $1.$ Subgame perfect and pre-commitment consumption rates for $\mu_{\bf{0}}=0.1, \mu_{\bf{1}}=0.1,
\sigma_{\bf{0}}=0.2, \sigma_{\bf{1}}=0.2,  r_{\bf{0}}=0.05,  r_{\bf{1}}=0.05;$  the discount rates $ \rho_{\bf{0}}=0.3,
\rho_{\bf{1}}=0.06.$


\noindent Fig. $2.$  Subgame perfect and pre-commitment consumption rates for $\mu_{\bf{0}}=0.1, \mu_{\bf{1}}=0.1,
\sigma_{\bf{0}}=0.2, \sigma_{\bf{1}}=0.2,  r_{\bf{0}}=0.01,  r_{\bf{1}}=0.09;$  the discount rates $ \rho_{\bf{0}}=0.07,
\rho_{\bf{1}}=0.06.$

 \begin{remark}
 Numerical findings reveal the complex dependence of the  subgame perfect and pre-commitment consumption rates
 on the model parameters. In figure $1$ we have the same market coefficients on the two states and a higher discount rate
 on the first state. This leads to higher  subgame perfect consumption rate in the first state (in
 figure $1$ notice that $\bar{C}(t,{\bf{1}})=\hat{C}_{2}(t,{\bf{0}})=\hat{C}_{2}(t,{\bf{1}}).$ In figure $2$ we also let the interest rate depend on the regime.
 Here although $$ \rho_{\bf{0}}=0.07> \rho_{\bf{1}}=0.06,$$
   $\bar{C}(t,{\bf{1}})>\bar{C}(t,{\bf{0}}).$ The difference between different consumption rates
 becomes more pronounced for higher $\gamma.$
 \end{remark}

\section{Appendix}

\subsection{A. Proof of Lemma \ref{l4.1} }
Existence and uniqueness for the ODE system is granted locally in time. For global
existence  and uniqueness we eastablish global estimates.
Let $\alpha(t,i)=-\left[\gamma r(t,i)
+\frac{\mu^2(t,i)\gamma}{2\sigma^2(t,i)(1-\gamma)}-\rho_i+\lambda_{ii}\right],  i\in \mathcal{S},$
and $\bar{\alpha}(t,i)=-\left[\gamma r(t,i)
+\frac{\mu^2(t,i)\gamma}{2\sigma^2(t,i)(1-\gamma)}-\rho_j+\lambda_{ii}\right],  i,j\in \mathcal{S}.$
Then from the ODE system we get
$$\frac{\partial g}{\partial t}(t,i)-\alpha(t,i){g}(t,i)\leq 0,\,\quad \frac{\partial \bar{g}}{\partial t}(t,i)-\bar{\alpha}(t,i){\bar{g}}(t,i)\leq 0,\,
 i\in \mathcal{S}.$$
Integrating this from $t$ to $T$ and using $g(T,i)=\bar{g}(T,i)=1,  i\in \mathcal{S}$ we get a global lower bound $M>0$
for $g,$ $\bar{g}.$ Next, since $\gamma<1,$ from the ODE system
$$\frac{\partial g}{\partial t}(t,i)\geq (\alpha(t,i)-(1-\gamma)M^{\frac{\gamma}{\gamma-1}}){g}(t,i)-\lambda_{ij}\bar{g}(t,j),  i,j\in \mathcal{S},$$
$$\frac{\partial \bar{g}}{\partial t}(t,j)\geq (\bar{\alpha}(t,j)-(1-\gamma)M^{\frac{\gamma}{\gamma-1}})\bar{g}(t,j)-\lambda_{ji}{g}(t,i),  i,j\in \mathcal{S}.$$
Let $A(t,i,j)\triangleq \min{(\alpha(t,i)-(1-\gamma)M^{\frac{\gamma}{\gamma-1}}, \bar{\alpha}(t,j)-(1-\gamma)M^{\frac{\gamma}{\gamma-1}})},$
$\lambda(i,j)\triangleq\max{( \lambda_{ij}, \lambda_{ji})}.$ Let $h(t)\triangleq g(t,i)+\bar{g}(t,j).$ Then by adding the above inequalities we get
$$\frac{\partial h}{\partial t}(t)\geq  (A(t,i,j)-\lambda(i,j)) h(t).$$ Integrating this from $t$ to $T$ yields the desired upper bound.

Recall that $v(t,x,i)=g(t,i)\frac{x^\gamma}{\gamma},$ and define $\bar{v}$ by
 $\bar{v}(t,x,i)\triangleq\bar{g}(t,i)\frac{x^\gamma}{\gamma}.$
 Then it can be easily shown that $(v, \bar{v})$ solves \eqref{zhang}, \eqref{zhang1}. Then by Lemma \ref{lala}, $v$ satisfies \eqref{00ie1}. 
\begin{flushright}
$\square$
\end{flushright}

\end{document}